\documentclass[pra,aps,twocolumn,preprintnumbers,amsmath,amssymb,superscriptaddress]{revtex4-1}
\usepackage{hyperref}
\usepackage{graphicx}

\begin{document}
\title{Dynamic Phase Diagram for the Quantum Phase Model}

\author{P. Buonsante}
\affiliation{QSTAR and INO-CNR, Largo Fermi 2, 50125 Firenze, Italia}
\author{L. Orefice}
\affiliation{QSTAR and INO-CNR, Largo Fermi 2, 50125 Firenze, Italia}
\affiliation{Dipartimento di Fisica, Universit\`a degli Studi di Bologna, Via Irnerio 46, 40126 Bologna, Italia}
\author{A. Smerzi}
\affiliation{QSTAR and INO-CNR, Largo Fermi 2, 50125 Firenze, Italia}

\begin{abstract}
We address  the stability of superfluid currents in a system of interacting lattice bosons. We consider various Gutzwiller trial states for the quantum phase model which provides a good approximation for the Bose-Hubbard model in the limit of large interactions and boson populations. We thoroughly analyze the current-carrying stationary states of the dynamics ensuing from a Gaussian ansatz, and derive analytical results for the critical lines signaling their modulational and energetic instability, as well as  the maximum of the carried current. We show that these analytical results are in good qualitative agreement with those obtained numerically in previous works on the Bose-Hubbard model, and in the present work for the quantum phase model. 
\end{abstract}
\maketitle

\section{Introduction}
Current-carrying stationary states of ultracold bosons in optical lattices are known to undergo a dynamical transition when the phase gradient associated with the flow exceeds a critical value. 
 In the regime of large densities and small interactions, where the discrete nonlinear Schr\"odinger
equations resulting from a tight-binding approximation of the Gross-Pitaevskii equation apply ,
 the critical phase gradient turns out to be $p=\frac{\pi}{2}$  \cite{Wu_PRA_64_061603R,Wu_NJP_5_104,Smerzi_PRL_89_170402,Smerzi_Chaos_13_766}. The inclusion of quantum fluctuations in the tight-binding description, via a Gutzwiller trial state, reveals that the critical value is a decreasing function of the effective interaction. Specifically, this function agrees with the  (discrete) Gross-Pitaevskii value for small effective interactions, and vanishes at the (mean-field) critical threshold for the superfluid-insulator quantum phase transition \cite{Altman_PRL_95_020402,Polkovnikov_PRA_71_063613}.
An experiment measuring the stability of the superfluid currents of ultracold atoms a three-dimensional optical lattice \cite{Mun_PRL_99_150604} provided results in remarkable agreement with the theoretical prediction \cite{Altman_PRL_95_020402,Polkovnikov_PRA_71_063613}. 
\vskip 2mm
Here we investigate the stability of  superfluid currents in the quantum phase model (QPM) that is known to be equivalent to the Bose-Hubbard model in the range of parameters where the latter exhibits its hallmark superfluid-insulator quantum phase transition \cite{vanOtterlo_PRB_52_16176,Huber_PRL_100_050404}.
The phase diagram for the stability of the superfluid current is first investigated in the mean-field picture derived from a factorized trial state whose factors have a Gaussian form depending on four dynamical variables. An analytic form for the modulational instability critical line is derived from the study of the linear dynamics of the perturbations over the current-carrying stationary states. The lines for one, two, and three-dimensional lattices are in good agreement with the results obtained for the Bose-Hubbard model from the integration of the mean-field dynamics \cite{Altman_PRL_95_020402,Polkovnikov_PRA_71_063613} and the numerical study of the excitation spectra \cite{Saito_PRA_86_023623}. 
Also, we are able to provide analytical results for the phase gradient attaining the maximal superfluid current, and for the energetic instability threshold. Both of these are found to coincide with the modulational instability threshold.
\vskip 2mm
We discuss some artifacts in the above picture, by comparing it against numerical  results for the most general product trial state. Finally, we introduce a new analytically tractable trial state, whose factors depend on a single parameter, and show that it provides a remarkably good approximation for some of the numerically obtained results.
\vskip 2mm
The plan of the paper is the following: In Section \ref{model} we briefly review the QPM and its relation with the Bose-Hubbard model; In Section \ref{anres} we introduce our Gaussian ansatz, and discuss the character of the current-carrying stationary states thereof; In Section \ref{numres} we compare  the analytical results derived in Section \ref{anres} against those obtained from a standard Gutzwiller description of the QPM, and provide arguments for ignoring some artifacts in the former; Section \ref{manres} contains some further analytical results based on a different product trial state, which do not suffer from the above artifacts and are in remarkable agreement with the numerical results obtained from the standard Gutzwiller ansatz discussed in \ref{numres}; Section \ref{conc} presents our conclusions. The most technical aspects of our analysis are confined to the Appendices.

\section{The Model}
\label{model}
The Bose-Hubbard model describes interacting bosons hopping across the sites of a lattice, and is characterized by a hallmark quantum phase transition between a superfluid and an insulating ground state. This is driven by the ratio of the interaction strength and the hopping amplitude and, on a translation-invariant lattice, it requires an integer average site occupancy  \cite{Fisher_PRB_40_546}.  Ultracold atoms trapped in the periodic potential formed by counterpropagating laser beams have been shown to provide an almost ideal experimental realization of such model \cite{Jaksch_PRL_81_3108,Greiner_Nature_415_39}. 

For large (integer) values of  the site occupancy and sufficiently strong effective interactions, the Bose-Hubbard model is equivalent \cite{vanOtterlo_PRB_52_16176,Huber_PRL_100_050404}  to the  simpler  quantum phase model \cite{Simanek_PRB_22_459}, whose Hamiltonian is 
\begin{equation}
\label{Hqpm}
\hat H = \frac{U}{2} \sum_\mathbf{r} \delta {\hat n}_\mathbf{r}^2 -K\sum_{\langle \mathbf{r}\,{\mathbf{r}'} \rangle} \cos (\hat \phi_\mathbf{r} - \hat \phi_{\mathbf{r}'}) 
\end{equation}
where we label a lattice site with the vector of its (discrete) coordinates. Specifically, $\mathbf{r} = \sum_{j=1}^{d} r_j \mathbf{e}_j $, where  $\mathbf{e}_j$ is the lattice vector along the $j-$th direction and  $r_j = 1, 2, \cdots, L_j$ is the corresponding coordinate.
The operators $\delta {\hat n}_\mathbf{r}$ and ${\hat \phi}_\mathbf{r}$ in Eq. \eqref{Hqpm} are conjugate, $[\hat \phi_\mathbf{r},\,\delta \hat n_{\mathbf{r}'}]=i\,\delta_{\mathbf{r} {\mathbf{r}'}}$, and describe  deviation from average occupancy and phase at site {\bf r}, respectively. The parameters $U$ and $K$ are the on-site interaction and ``Josephson coupling", respectively, the latter being related to  the average occupancy , $\bar n$, and hopping amplitude of the underlying Bose-Hubbard model,  $J$, as $K = 2 J \bar n$. 


In the following sections we will assume that the system is described by a factorized trial state of the form
\begin{equation}
\label{prodTS}
|\Psi\rangle = e^{i {\cal S}(t)} \prod_\mathbf{r} |\psi_\mathbf{r}\rangle
\end{equation}
where each of the factors refers to a lattice site \footnote{Throughout this paper we set $\hbar=1$. The role of the time-dependent overall phase factor ${\cal S}(t)$ will be clarified in Appendix \ref{Atdvp}}.
Spatially uniform stationary states characterized by a constant phase gradient in the ``local order parameter",
\begin{equation}
\alpha_\mathbf{r}(p)=\langle \Psi | e^{i \hat \phi_{\mathbf{r}}}|\Psi\rangle = e^{i p r_j} \alpha(p)
\end{equation}
carry a current 
\begin{align}
\label{cur_mf}
{\cal J}(p) &=2 K |\langle \Psi | \sin\left( \hat \phi_{\mathbf{r}+\mathbf{e}_j}- \hat \phi_{\mathbf{r}}\right)|\Psi\rangle |^2 \nonumber\\
&=2 K |\alpha(p)|^2 \sin p
\end{align}
along coordinate direction $\mathbf{e}_j$. The (site-independent) square modulus of the local order parameter can be therefore identified with the superfluid density $\rho_{\rm S} = |\alpha(p)|^2$ \cite{Polkovnikov_PRA_71_063613}.  

 As we discuss in the following,  the value of $|\alpha(p)|$ for a current-carrying state with momentum $p$ at a given value of the hopping amplitude is the same as that for the ground-state, $|\alpha(0)|$, at a rescaled value of the hopping amplitude, $K'=K (d-1+\cos p)$. This is a general property of uniform stationary states of the form in Eq. \eqref{prodTS}, applying e.g. also in the Gutzwiller approach to the Bose-Hubbard model.
\section{Analytical results}
\label{anres}
In the present section we assume that the factors in Eq.~\eqref{prodTS}, 
 have a Gaussian form depending on four dynamical variables \footnote{Ref. \cite{Huber_PRL_100_050404} makes use of an equivalent ansatz to analyze the amplitude and phase modes in the ground state of the quantum phase model.}
\begin{equation}
\label{ansatzG}
|\psi_\mathbf{r}\rangle = \frac{1}{\sqrt[4]{4 \pi \sigma_\mathbf{r}}} e^{-\left(\frac{1}{8\sigma_\mathbf{r}}-i \frac{\varepsilon_\mathbf{r}}{2}\right)(\phi_\mathbf{r}-\varphi_\mathbf{r})^2+i \kappa_\mathbf{r}(\phi_\mathbf{r}-\varphi_\mathbf{r})}
\end{equation}
A standard procedure \cite{Jackiw_PLA_71_158}, described in more detail inAppendix \ref{Atdvp}, provides the equations of motion for the above dynamical variables, 
\begin{equation}
\label{eqmot}
\left\{\begin{array}{l}
\dot \varphi_\mathbf{r} = U \kappa_\mathbf{r} \\
\dot \kappa_\mathbf{r} = K e^{-\sigma_\mathbf{r}} {\sum_{\mathbf{r}'}'} e^{-\sigma_{\mathbf{r}'}} \sin (\varphi_{\mathbf{r}'} - \varphi_\mathbf{r}) \\
\dot \sigma_\mathbf{r} = 2U \sigma_\mathbf{r} \varepsilon_\mathbf{r} \\
\dot \varepsilon_\mathbf{r} = \frac{U}{16\sigma_\mathbf{r}^2}\!-\! U\varepsilon_\mathbf{r}^2\!-\! K e^{-\sigma_\mathbf{r}} {\sum_{\mathbf{r}'}'} e^{-\sigma_{\mathbf{r}'}} \cos (\varphi_{\mathbf{r}'} - \varphi_\mathbf{r})
\end{array}
\right.
\end{equation}
where the prime signals that the sum is restricted to the sites $\mathbf{r}'$ adjacent to $\mathbf{r}$.
Equations \eqref{eqmot} ensue from the semiclassical Hamiltonian
\begin{align}
{\cal H} &= \sum_\mathbf{r} \frac{U}{2}\left(\kappa_\mathbf{r}^2+\frac{1}{8\sigma_\mathbf{r}}+2\sigma_\mathbf{r} \varepsilon_\mathbf{r}^2\right) \nonumber\\
\label{mfHG}
& - K \sum_{\langle  \mathbf{r}\, {\mathbf{r}'} \rangle} e^{-\sigma_\mathbf{r} -\sigma_{\mathbf{r}'}}\cos(\varphi_\mathbf{r}-\varphi_{\mathbf{r}'})
\end{align}
equipped with the Poisson brackets
\begin{equation}
\{\varphi_\mathbf{r},\kappa_{\mathbf{r}'}\} = \delta_{\mathbf{r}\,{\mathbf{r}'}}, \qquad \{\sigma_\mathbf{r},\varepsilon_{\mathbf{r}'}\} = \delta_{\mathbf{r}\,{\mathbf{r}'}}.
\end{equation}
It is easy to check that the choice
\begin{equation}
\label{ccss}
\varphi_\mathbf{r} = p\, r_j,\quad \kappa_\mathbf{r} = 0,\quad \sigma_\mathbf{r} = \bar \sigma,\quad \varepsilon_\mathbf{r} = 0
\end{equation}
corresponds to a stationary state 
carrying a current  
\begin{equation}
\label{ccss_cur}
{\cal J}(p) = 2 K e^{-2 \bar \sigma} \sin p
\end{equation}
along  direction $\mathbf{e}_j$, whose energy and superfluid density  are
\begin{equation}
\label{nrg0}
E_p =2 d  K \left[\frac{\Gamma}{16 \sigma}  - e^{-2\,\bar \sigma} \frac{d-1+\cos p}{2 d} \right],\quad \rho_{\rm S} = e^{-2 \sigma}
\end{equation}
where we introduced the effective interaction parameter $\Gamma = \frac{U}{2 d K}$.
The parameters $p$ and $\bar \sigma$ in Equations \eqref{ccss}--\eqref{nrg0} must obey the following relations
\begin{align}
\label{ccss_p1}
p \frac{L_x}{2\pi} &= 0,1,\ldots,L_x-1,\\
\label{ccss_p2}
\bar \sigma(\Gamma,p) &= -{\cal W}\left(-\sqrt{\frac{\Gamma \,d}{16 (d-1+\cos p)}}\right)
\end{align}
where $L_x$ is the linear size of the lattice in the $x$ direction, $d$ is the dimensionality of the lattice and  ${\cal W}(x)$ is the so-called {\it Lambert W function} which provides the solution to  the equation
\begin{equation}
\label{Gamma}
\Gamma = 16\, \bar \sigma^2 e^{-2 \bar \sigma} \frac{d-1+\cos p}{d}
\end{equation} 
This is obtained by plugging Eq. \eqref{ccss} in the last Eq. \eqref{eqmot}, and provides a relation between $p$, $\bar \sigma$ and $\Gamma$. Also notice that Eq. \eqref{Gamma} can be obtained from the stationarization of the energy density in Eq. \eqref{nrg0} with respect to the Gaussian width $\bar \sigma$. 

According to Eq. \eqref{ccss_p2}, the finite $\bar \sigma$ necessary for a nonvanishing current  only exists below a $p-$dependent threshold 
\begin{equation}
\label{crpt1}
\Gamma_{p }^{(1)} = 16 \,e^{-2} \frac{d-1+\cos p}{d}
\end{equation}
Note that, since $\Gamma \geq 0$, on one dimensional lattices it should be $|p| < \frac{\pi}{2}$.
Also note that for $0 < \Gamma < \Gamma_{p }^{(1)}$ Eq. \eqref{Gamma} admits two solution, corresponding to the two main branches ${\cal W}_0$ and ${\cal W}_1$ of the Lambert function in Eq. \eqref{ccss_p2}. 
At vanishing momentum the threshold in \eqref{crpt1}, $\Gamma_0^{(1)} = 16 \,e^{-2}$, can be interpreted as the critical point between a superfluid ground state, $0< \bar \sigma < \infty$, and an insulating state, $\bar \sigma = \infty$. The latter choice in fact corresponds to a trivial stationary solution of Eq. \eqref{eqmot} irrespective of $\Gamma$ and $p$, as it can be  checked by direct substitution. The insulating character of this solution is apparent from the vanishing of the superfluid density and current,  Eqs. \eqref{ccss_cur} and  \eqref{nrg0} (see Section \ref{numres} for more detail).

Of course one can easily write a stationary state whose current flows along a different direction, possibly not parallel to a coordinate axis. In the following we are concerned  with the properties of current-carrying states on one-, two- and three-dimensional lattices. 

\begin{figure}
\begin{center}
\includegraphics[width=8cm]{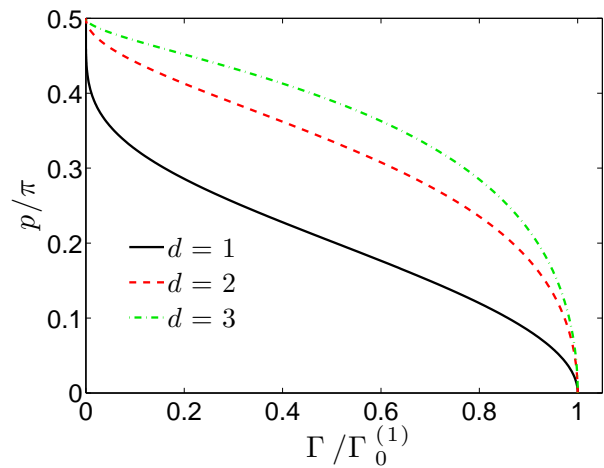}
\end{center}
\caption{\label{Fmodinst} (color online) Threshold for the dynamical instability of stationary states carrying a current along one of the coordinate directions of the lattice, as provided by Eq. \eqref{Gmodinst}. 
The effective interaction in the abscissae is normalized by the upper bound for the existence of a nontrivial solution at $p=0$, Eq. \eqref{crpt1}, which can be seen as the critical point for the superfluid to insulator quantum phase transition in the quantum phase model. 
Note the qualitative agreement of the curves in this figure with the analogous results obtained numerically in Refs. \cite{Altman_PRL_95_020402,Polkovnikov_PRA_71_063613,Saito_PRA_86_023623}.  As discussed in Section  \ref{anres}, the same lines also identify the stationary states attaining the maximum current, and signal the threshold for energy instability. }
\end{figure}

According to a standard procedure, briefly reviewed in Appendix \ref{BogFreq}, the dynamic stability of the current-carrying states can be inferred from the spectrum of the matrix governing the (linear) dynamics of the small perturbations.  This matrix can be conveniently analyzed in the reciprocal lattice where, owing to the translation invariance of the system, it has a block diagonal form.
Each of its  $4\times 4$ blocks is labeled by the reciprocal lattice label $\mathbf{q}$, and its fourth-degree characteristic polynomial turns out to have the simple form
\begin{equation}
\label{char_pol}
\omega^4 + 2 b(p,\mathbf{q})\, \omega^2 + c(p,\mathbf{q}) 
\end{equation}
where the explicit form of the coefficients is given in  Appendix \ref{BogFreq}.
Its roots  $\omega_{j}(p,\mathbf{q})$ are the  {\it Bogoliubov frequencies} of the stationary state,  \footnote{Actually only half of the frequencies, usually the positive ones, must be taken into account, the remaining ones being redundant.}. 
A stationary state characterized by a given $p$ is dynamically (or {\it modulationally}) stable when the latter are real for all $j$'s and $\mathbf q$, i.e. when 
\begin{equation}
\label{stability}
b^2(p,\mathbf{q})- c(p,\mathbf{q})>0, \quad b(p,\mathbf{q}) < 0, \quad c(p,\mathbf{q}) > 0
 \end{equation}
Since $\bar \sigma > 1$ when Eq. \eqref{ccss_p2} refers to the ``$-1$" branch of the Lambert function  $c(p,\mathbf{q})$ is always negative, as discussed in Appendix \ref{BogFreq}. This is in agreement with the fact that the energy density of the stationary state, Eq. \eqref{nrg0}, corresponds to a local maximum. Conversely $\bar \sigma <1$ on the ``$0$" branch of the Lambert function, and the conditions in Eq. \eqref{stability}  add up to 
\begin{equation}
\label{Fd}
\bar \sigma < \frac{1}{2} F_d(p),\qquad F_d(p) = 2 \cos p \frac{d-1+\cos p}{1+(d-1)\cos p}
\end{equation}
with $p \in \left(0,\,\frac{\pi}{2}\right)$. Plugging this condition into Eq. \eqref{Gamma} results in the dynamical instability threshold 
\begin{equation}
\label{Gmodinst}
\Gamma_{p}^{(2)} = 4 \frac{d-1+\cos p}{d}\,F_d^2(p)\, e^{-F_d(p)}
\end{equation}
which we plot in
Fig. \ref{Fmodinst} for one--, two-- and three--dimensional lattices. These analytic results are qualitatively similar to those obtained numerically in Refs. \cite{Altman_PRL_95_020402,Polkovnikov_PRA_71_063613,Saito_PRA_86_023623} for the Bose-Hubbard model. We note that for vanishing $p$ the threshold for dynamical instability collapses to the critical point for the transition to the insulating state, $\Gamma^{(2)}_0=\Gamma^{(1)}_0$, which  is also in agreement with the mentioned references. 

In Refs.  \cite{Altman_PRL_95_020402,Polkovnikov_PRA_71_063613} the modulational instability threshold is linked to the maximum possible current flowing through the system. Our ansatz, Eq. \eqref{ansatzG}, allows us to work out analytically the function $\Gamma^{(3)}_p$ giving the relation between the effective interaction and   value of $p$ attaining the maximum  current, and  and directly verify that it coincides with the function in Eq. \eqref{Gmodinst}. 

 The requirement that the current, Eq. \eqref{ccss_cur}, has a maximum results in a relation between $\bar \sigma$ and $p$ which, plugged in Eq. \eqref{Gamma} produces the function $\Gamma^{(3)}_p$ we are looking for.
The study of the maximum of Eq. \eqref{ccss_cur} requires an explicit expression for the derivative of $\bar \sigma$ with respect of $p$. From the properties
of the Lambert function or, equivalently,  by deriving Eq. \eqref{Gamma} at fixed $\Gamma$
we obtain
\begin{equation}
\frac{d \bar \sigma}{d p} =\frac{\bar \sigma}{1-\bar \sigma} \frac{\sin p}{2 (d-1+\cos p)}
\end{equation} 
and hence
\begin{equation}
\label{dcurr}
\frac{d {\cal J}}{d p} = 2 K e^{-2 \bar\sigma}\frac{\cos p}{1-\sigma}\left[1-\frac{2\bar \sigma}{F_{\rm d}(p)} \right]
\end{equation}
When Eq. \eqref{ccss_p2} refers to the main branch of the Lambert function one gets  $0 \leq \bar \sigma \leq 1$, and --  irrespective of the dimension -- the current has always a maximum at a $p$ in $\left[-\frac{\pi}{2}, \frac{\pi}{2} \right]$ corresponding to the condition $2 \bar \sigma = F_d(p)$ where the quantity in the square brackets of Eq. \eqref{dcurr} vanishes. This is in agreement with threshold in Eq. \eqref{Fd}, obtained from the study of the Bogoliubov frequencies.
As to the  ${\cal W}_{-1}$ branch of the Lambert function, one gets  $1 \leq \bar \sigma < \infty$.
On one- and two-dimensional lattices the current is an increasing function of $p$, whereas if $d\geq 2$ it exhibits once again a maximum for $2 \bar \sigma = F_d(p)$, which in this case corresponds to 
$|p|>\frac{\pi}{2}$. However in the present case  the derivative of the current has no relation with the  character of the stationary state, which is always unstable,  as discussed  in  Appendix \ref{BogFreq}.
This has  to do with the fact that the stationary states with $\bar \sigma>1$ correspond to maxima of the energy density, Eq. \eqref{nrg0}, and are hence energetically unstable. 

Energy instability occurs when a small perturbation is able to lower the system energy \cite{Wu_NJP_5_104,Menotti_NJP_5_112,DeSarlo_PRA_72_013605} and, in general is a necessary but not sufficient condition for modulational instability. For the quantum phase model the thresholds for these two different kinds of instability turn out to coincide. The threshold for energetic instability can be equivalently found by requiring that either the Bogoliubov spectrum \cite{Menotti_NJP_5_112} or  the energy spectrum \cite{Wu_NJP_5_104}  contain at least one vanishing element.
According to the above discussion,   the sufficient condition for a vanishing Bogoliubov frequency is that the coefficient $c(p,\mathbf{q})$ appearing in Eq. \eqref{char_pol} and explicitly defined in Eq. \eqref{cp_c} vanishes. But, as discussed in  Appendix \ref{BogFreq} this once again results in the function $\Gamma^{(2)}_p$ giving the modulational instability threshold.
 The same result can be obtained by studying the energy spectrum for the small perturbations on the stationary state \cite{Wu_NJP_5_104}.  As discussed in some detail in  Appendix \ref{EnFreq},  once again this is obtained from a matrix which, owing to the translation invariance, reduces to a collection of $4\times 4$ independent blocks labeled by $\bf q$, whose characteristic polynomial is 
\begin{equation}
\label{char_pol_E}
(\lambda -d\,\Gamma)(\lambda - 2 d\, \Gamma \bar \sigma) \left[\lambda^2 + 2\,b'(p,{\bf q}) \, \lambda +c'(p,{\bf q})\right]
\end{equation}
where the coefficients of the quadratic factor are explicitly given in  Appendix \ref{EnFreq}. Since $\Gamma$ and $\bar \sigma$ are always positive, the first two linear factors correspond to positive eigenvalues. Therefore the critical conditions for energy stability has once again the form in Eq. \eqref{stability}. As discussed in more detail in  Appendix \ref{EnFreq}, the stationary states with $\bar \sigma >1$ are always energetically unstable, since for any $p>0$ there exist some $\mathbf{q}$ such that $b'(p,\mathbf{q})>0$. Conversely, for $0<\bar \sigma < 1$, $b'$ is always negative, so that the stability character is determined by $c'(p,\mathbf{q})$. Since this has the same sign as $c(p,\mathbf{q})$ (the two quantities differ by a positive constant), it is clear that  the threshold for energetic instability is exactly the same as that for modulational instability. As we mention, this result does not hold true in general. Studies based on the Gross-Pitaevskii equation \cite{Wu_NJP_5_104}, discrete nonlinear Schrodinger equation \cite{Menotti_NJP_5_112} and the Gutzwiller equations for the Bose-Hubbard model \cite{Saito_PRA_86_023623,Gutz_inprep} show that there exists a finite range of $p$ in which the stationary states are dynamically stable but energetically unstable. However, it can be shown \cite{Gutz_inprep} that
this interval shrinks with increasing interaction and average site occupancy, and it is hence expected to be negligible in the regime where the quantum phase model applies.

%
%

\section{Comparison against numerical results}
\label{numres}
In the present section we compare the analytical results obtained in the previous section against those obtained numerically from a standard mean-field approach to the quantum phase model.
Assuming a factorized trial state of the form in Eq. \eqref{prodTS} with no further constraint on the factors, one is left with a set of on-site Hamiltonians
\begin{equation}
\label{mfHgen}
{\cal H}_{\mathbf r} =  \frac{U}{2} \delta {\hat n}_\mathbf{r}^2 -\frac{K}{2} \left(\eta_\mathbf{r}^*\, e^{i \hat \phi_{\mathbf r}}+\eta_\mathbf{r}\, e^{-i \hat \phi_{\mathbf r}} \right),\end{equation}
where the complex parameter $\eta_\mathbf{r}$ pertaining to site $\mathbf{r}$ depends on the local order parameters at the neighboring sites
\begin{equation}
\label{mfscgen}
\eta_\mathbf{r} = {\sum_{\mathbf{r}'}}' \alpha_{\mathbf{r}'}
\end{equation} 
and must be determined self-consistently.
A solution corresponding to uniform stationary state carrying a current along  coordinate direction $\mathbf{e}_j$  is obtained by setting
\begin{equation}
\label{TSfactorE}
|\psi_{\mathbf r}\rangle = \sum_{\nu = -\infty}^{\infty} e^{i \nu p r_j }f_{\nu} |\nu \rangle_{\mathbf r},\quad  \sum_{\nu = -\infty}^{\infty} \left|f_{\mathbf r}^{(\nu)}\right|^2 = 1
\end{equation}
where 
\begin{equation}
|\nu\rangle_{\mathbf r} = \frac{1}{\sqrt{2 \pi}} e^{i \nu \phi_{\mathbf r}}
\end{equation}
is an eigenstate of the number fluctuations $\delta \hat n_{\mathbf r}$. The complete determination of the stationary state is thus reduced to the calculation of the site-independent coefficients $f_\nu$, which we assume to be real without loss of generality. These are the solutions of the single-site self-consistent problem
\begin{align}
\label{mfHss1}
{\cal H} &=  \frac{U}{2} \delta {\hat n}^2 -2 K (d-1+\cos p) \alpha \cos  \hat \phi \\
\label{mfHss2}
\alpha & = \langle \psi | e^{i \hat \phi}|\psi\rangle = \langle \psi | \cos \hat \phi \, |\psi\rangle, \quad  |\psi \rangle = \sum_{\nu = -\infty}^{\infty}f_{\nu} |\nu \rangle
\end{align}
where we dropped the site label $\mathbf r$. Thus, as we mention in Sec.~\ref{model}, the problem of finding a current-carrying stationary state is formally equivalent to finding the ground state  of the system , yet with a rescaled hopping amplitude $K\; \to\; \tilde K =K (d - 1 + \cos p)/d $ \cite{Gutz_inprep}.  
The ground-state for the $d$-dimensional system was discussed e.g. in Ref. \cite{Simanek_PRB_22_459}, were it was found that the superfluid-insulator transition occurs at $U/d \tilde K =4$. Thus, a  stationary state carrying a current in one coordinate direction is found for effective interactions below 
\begin{equation}
\label{crpt_num}
\bar \Gamma^{(1)}_p = 2  \frac{d-1+\cos p}{d}.
\end{equation} 
Conversely, for $\Gamma>\Gamma^{(1)}_p$, only insulating stationary states are found, $\alpha=0$. Note that the threshold in Eq. \eqref{crpt1} differs from the ``exact one", Eq. \eqref{crpt_num} by a mere $8\%$. However the simplified mean-field treatment of Section \ref{anres} suffers from some artifacts. First of all, unlike $\alpha^2$ in the present "exact" treatment, the order parameter $e^{-2 \bar \sigma}$ does not vanish at the critical value above which only solutions with $\bar \sigma = \infty$ are found, Eq \eqref{crpt1}. Note indeed that, according to Eq. \eqref{ccss_p1}, 
\begin{equation}
\label{critsigma}
\bar \sigma\left(\Gamma^{(1)}_p,p\right) = -{\cal W}\left(-\frac{1}{e}\right) = 1  
\end{equation} 

Also, there exists an interval of effective interactions $\Gamma_p^{(4)} <  \Gamma < \Gamma_p^{(1)}$, with
\begin{equation}
\label{crpts1}
\Gamma_p^{(4)} = 4 e^{-1} (d-1+\cos p),
\end{equation}
in which the stationary states featuring a finite $\bar \sigma$ which satisfies Eq. \eqref{ccss_p2} have a higher energy than that pertaining to the insulating state, $\bar \sigma = \infty$. 
From a slightly different perspective,  the latter always corresponds to a  minimum of the Ginzburg-Landau functional, which becomes the absolute minimum for $\Gamma > \Gamma_p^{(4)}$.
Thus, energy arguments would have  $\Gamma_0^{(4)}$ as the critical point for the superfluid-insulator transition in the mean-field treatment of Section \ref{numres}. Once again, the order parameter would not only be finite at the phase transition, but it rather would be even larger than at $\Gamma^{(1)}_p$. Indeed
\begin{equation}
\label{crsp}
\bar \sigma\left(\Gamma^{(4)}_p,p\right) = -{\cal W}\left(-\frac{1}{2 e}\right) < 1  
\end{equation} 
\begin{figure}
\begin{center}
\includegraphics[width=8cm]{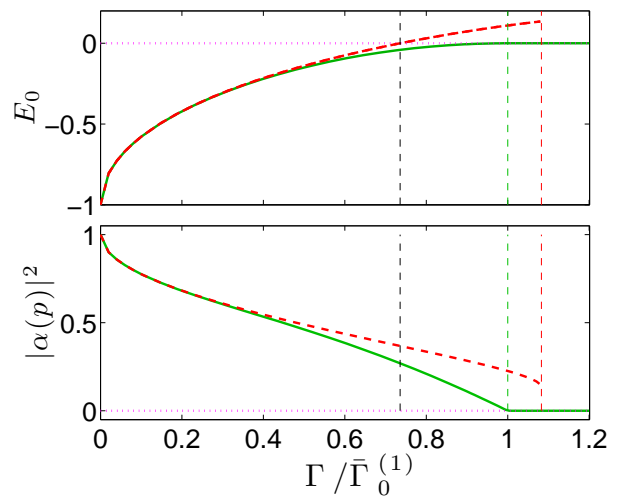}
\end{center}
\caption{\label{Fartifacts} (color online)  Comparison between the mean-field approaches of Sections \ref{anres} and \ref{numres}. The top and bottom panels show the behavior of the energy density and order parameter, respectively. The thick green solid lines refer to the numerical results obtained in Sec. \ref{numres}. The thick red dashed lines are the corresponding analytic results, Eq. \eqref{nrg0} and $e^{-2\bar \sigma}$, with $\bar \sigma$ given by Eq. \eqref{ccss_p2}. The dotted magenta horizontal lines  refer to the insulating state. The red and green dashed vertical lines signal the critical point for the transition as provided by Eqs. \eqref{crpt1} and \eqref{crpt_num}, respectively. The dashed black vertical line signals the spurious critical point, Eq. \eqref{crsp}. }
\end{figure}
Figure. \ref{Fartifacts} illustrates the above-discussed artifacts, comparing  energy and order parameter for the quantum phase transition as provided by the "exact" mean field treatment with those obtained analytically in the simplified scheme of Sec. \ref{anres}.

\begin{figure}
\begin{center}
\includegraphics[width=8cm]{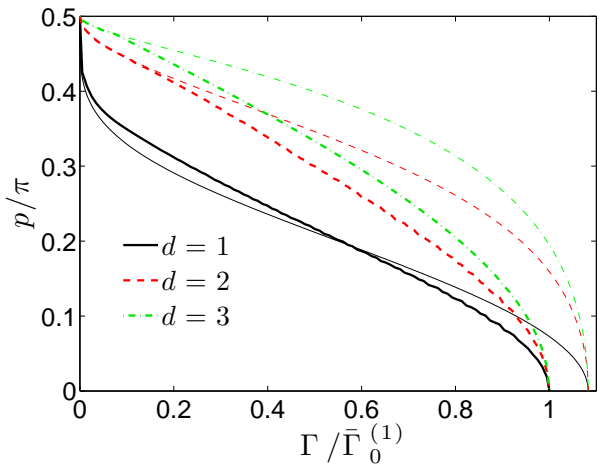}
\end{center}
\caption{\label{FmodinstEx} (color online) Maximum of the current, Eq. \eqref{cur_mf}, as obtained numerically from the self-consistent problem in Eqs. \eqref{mfHgen} and \eqref{mfscgen} (thick lines). The critical point used as the scale in the abscissae, is $\bar \Gamma_0^{(1)}= 2 $, as provided from the ``exact" mean-field analysis, Eq. \eqref{crpt_num}. We include the curves shown in Fig. \ref{Fmodinst} (thin lines) for comparison.}
\end{figure}

Several facts suggest that the critical point in Eq. \eqref{crpts1} should be discarded as spurious. First of all, as we mention, the critical point in Eq. \eqref{crpt1} is closer than Eq. \eqref{crpts1} to the ``exact'' value in Eq. \eqref{crpt_num}. Also, the critical lines in Fig. \ref{Fmodinst} are qualitatively similar to the results for the Bose-Hubbard  model reported in Refs. \cite{Altman_PRL_95_020402,Polkovnikov_PRA_71_063613,Saito_PRA_86_023623}, as well as to those in the present Fig.  \ref{FmodinstEx}, signaling the maximum of the current
carried by a stationary state described by Eqs. \eqref{prodTS} and \eqref{TSfactorE}, i.e. of Eq.~\eqref{cur_mf} where the local order parameter is determined from Eqs.~\eqref{mfHss1} and \eqref{mfHss2}.
If the transition to the insulating state occurred at $\Gamma^{(4)}_p$, the critical lines in Fig. \ref{Fmodinst} would have the same shape, but would not tend to the critical point of the quantum phase transition for arbitrarily small currents. This would be at variance with what observed in Refs. \cite{Altman_PRL_95_020402,Polkovnikov_PRA_71_063613} for the Bose-Hubbard model, and in Fig.  \ref{FmodinstEx} for the quantum phase model.

A further reason for identifying $\Gamma^{(1)}_0$ with the critical point comes from Ref. \cite{Huber_PRL_100_050404} where a description equivalent to that provided by Eqs. \eqref{prodTS} and \eqref{ccss} is employed in the investigation of the phase (Goldstone) and amplitude (Higgs) modes in the collective excitations over the ground state of the quantum phase model. These correspond to the two positive branches 
\begin{equation}
\omega_{1,2}(p,{\mathbf q}) = \sqrt{b(p,{\mathbf q})\pm \sqrt{b^{2}(p,{\mathbf q})-c(p,{\mathbf q})}}
\end{equation}
 of the Bogoliubov spectrum obtained from Eq. \eqref{char_pol}, in the case $p=0$, and the vanishing of the gap between them for  $\mathbf{q}\;\to\;0$ at $\Gamma=\Gamma^{(1)}_0$ is associated to criticality. 
 Note that on the ground-state, $p=0$, the $\bf q$-dependent $4\times4$ matrices giving the polynomials in Eq. \eqref{char_pol} further decouple into two $2\times 2$ independent blocks, so that  the two branches of the Bogoliubov spectrum can be naturally ascribed to phase ($\varphi_{\mathbf r}$ and $\kappa_{\mathbf r}$) and amplitude ($\sigma_{\mathbf r}$ and $\varepsilon_{\mathbf r}$) variables \cite{Huber_PRL_100_050404}. While this does not apply for the excited states, $p\neq 0$, where a mixing of phase and amplitude variables occurs, it is still true that the gap between the two Bogoliubov branches closes at $\Gamma=\Gamma^{(1)}_p$ for vanishing $\mathbf{q}$. Indeed, as it is clear from Eqs. \eqref{cp_c}-\eqref{C2} in  Appendix \ref{BogFreq}, the coefficient $c(p,\mathbf{q})$ vanishes under these circumstances.

\section{An improved trial state}
\label{manres}
In the previous section we compared the analytic results of the Gaussian mean-field ansatz of Sec. \ref{anres} with those obtained numerically by using a more general trial state, and discussed some artifacts of the former picture. 

One might be tempted to ascribe these artifacts to the normalization choice adopted in Eq. \eqref{ansatzG}. Note indeed that 
\begin{equation}
\langle \psi_\mathbf{r}|\psi_\mathbf{r} \rangle = \frac{1}{2 \sqrt{ \pi \bar \sigma}}\int_{-\bar \phi}^{-\bar \phi} d\phi_\mathbf{ r} \,e^{-\frac{(\phi_\mathbf{ r}-\varphi_\mathbf{ r})^2}{4 \bar \sigma}} 
\end{equation}
equals $1$ only if $\bar \phi = \infty$. As a matter of fact, $\bar \phi = \pi$, so that the chosen trial state is correctly normalized only for sufficiently small $\bar \sigma$.
\begin{figure}
\begin{center}
\includegraphics[width=8cm]{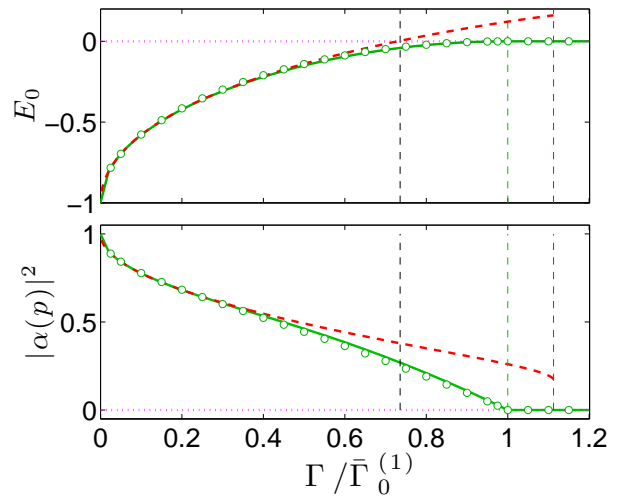}
\end{center}
\caption{\label{Fartifacts2} (color online) Comparison between the mean-field approaches of Sections \ref{numres} and \ref{manres}. The top and bottom panels show the behavior of the energy density and order parameter, respectively. The thick green solid lines refer to the numerical results obtained in Sec. \ref{numres}. The thick red dashed lines are the corresponding analytic results obtained from the minimization of Eq. \eqref{gsnrgG2}. The dotted magenta horizontal lines  refer to the insulating state. The  green dashed vertical line once again signals the critical point given by Eq. \eqref{crpt_num}, while the red dashed vertical line signals the critical point obtained numerically from the minimization of Eq. \eqref{gsnrgG2}, $\Gamma \approx 2.225 $. The dashed black vertical line signals the spurious critical point where the energy of the ``Gaussian" superfluid and insulating state are the same. The green circles have been obtained from the minimization of Eq. \eqref{gsnrgV}. }
\end{figure}
A possibly more effective ansatz for the ground state would be
\begin{equation}
\label{normal}
|\psi_\mathbf{r} \rangle = \frac{1}{\sqrt{\sqrt{4 \pi \bar \sigma}\,{\rm erf}\left(\frac{\pi}{2 \sqrt{\bar \sigma}}\right)}}e^{-\frac{\phi_\mathbf{ r}^2}{8 \bar \sigma}} 
\end{equation}
which would correspond to an energy density
\begin{equation}
\label{gsnrgG2}
E_0 = \frac{U}{8 \bar \sigma} \left[1+ \frac{\sqrt{\pi}}{\bar \sigma^2\,{\rm erf}\left(\frac{\pi}{2 \sqrt{\bar \sigma}}\right)} e^{-\frac{\phi_\mathbf{ r}^2}{4 \bar \sigma}} \right]- 2 d K \alpha^2
\end{equation}
where the order parameter is
\begin{equation}
\alpha = e^{-\bar \sigma} \frac{{\rm erf}\left(\frac{\pi+2\,i\sigma}{2 \sqrt{\bar \sigma}}\right)+{\rm erf}\left(\frac{\pi-2\,i\sigma}{2 \sqrt{\bar \sigma}}\right)}{2\,{\rm erf}\left(\frac{\pi}{2 \sqrt{\bar \sigma}}\right)}.
\end{equation}
Minimizing the functional in Eq. \eqref{gsnrgG2} gives the red dashed lines in shown in Fig.  \ref{Fartifacts2}. The comparison with Fig.  \ref{Fartifacts} shows that  the ``correct" normalization does not remove, but in fact worsens the artifacts discussed in Sec. \ref{numres}. Note indeed that the spurious critical point is basically unaffected by the normalization, while the transition point
moves to $\Gamma \approx 2.225$, farther from the ``exact" mean-field value $\Gamma = 2$ than $\Gamma_0^{(1)} = 16 e^{-2} \approx 2.165$. Also, the transition remains of the first order.
As it will be clear shortly, the above artifacts are related to the smoothness of the trial state. Note indeed that the factor in Eq. \eqref{normal} is of class $C^{0}$, its derivatives having different limits at the boundaries of the interval $(-\pi,\pi)$.

A one-parameter function of class $C^\infty$ which, similar to Eq. \eqref{gsnrgG2}, interpolates between a Dirac delta and a constant, is provided by
 \begin{equation}
 \label{Nansatz}
|\psi_\mathbf{r} \rangle = \frac{1}{\sqrt{I_0(2\gamma)}}e^{\gamma \, \cos \left( \hat \phi_\mathbf{r} - p r_j \right)}, 
\end{equation}
where $I_\alpha(x)$ denotes a modified Bessel function of the first kind, and we allowed for a phase difference in the $\mathbf{e}_j$ direction \footnote{It can be shown that Eq. \eqref{Nansatz} is actually a coherent state of the $E(2)$ algebra in its (operator) realization generated by $\cos \phi$, $\sin \phi$ and $-i \frac{\partial}{\partial \phi}$ \cite{Carruthers_RMP_40_411,Kastrup_PRA_73_052104}.}.  The corresponding energy density is 
\begin{equation}
\label{gsnrgV}
E_p = \alpha\!  \left[\frac{U\,\gamma}{4}-K\, (d-1+\cos p) \,\alpha\right], \,\,\; \alpha = \frac{I_1(2\,\gamma)}{I_0(2\,\gamma)}
\end{equation}
For $\gamma=0$ the present product trial state  corresponds to an insulating state. Indeed, as it is clear from   Eq. \eqref{Nansatz}, the phase is entirely undetermined, being the phase distribution constant. Also, the order parameter consistently vanishes, since $I_1(0)=0$. 
Conversely, $\gamma>0$ corresponds to a superfluid state, with $\alpha>0$. The critical point for the transition can be worked out analytically by treating $\gamma$ as a perturbative quantity. For small $\gamma$ one has $\alpha = \gamma$ and
\begin{equation}
E_p \approx   \left[\frac{U}{4}-K\, (d-1+\cos p) \right] \, \gamma^2
\end{equation}
which entails that $\gamma=0$ is the absolute minimum --- i.e. that the system is in an insulating state ---for $\Gamma$ exceeding the threshold in Eq. \eqref{crpt_num}. We remark that the above derivation is perturbative, but the resulting critical point is exact. Thus the one-parameter trial state in Eq. \eqref{Nansatz} yields the same critical point as the ``exact" mean-field treatment in Sec. \ref{numres}  \cite{Simanek_PRB_22_459}.

\begin{figure}
\begin{center}
\includegraphics[width=8cm]{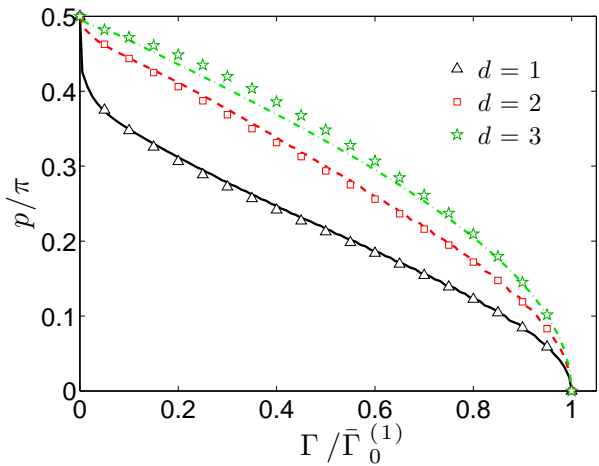}
\end{center}
\caption{\label{FmodinstEx2} (color online) Critical lines for the maximum of the current. The symbols refer to the results obtained from the minimization of the energy in Eq. \eqref{gsnrgV}. The ``exact" curves of Fig. \ref{FmodinstEx} are included for comparison. }
\end{figure}

More in general, the energy density and order parameters obtained from the  minimization of  Eq. \eqref{gsnrgV} with respect to $\gamma$ at fixed $\Gamma$ and $p=0$ (ground-state) show remarkable agreement with the numerical results obtained in Sec. \ref{numres}, as it is clear from Fig. \ref{Fartifacts2}.
The minimization of Eq. \eqref{gsnrgV} can be also used to calculate the maximum of the current
carried by an excited state, $p\neq 0$. The relevant equation is once more Eq. \eqref{cur_mf}, with $\alpha$ given by the minimization of Eq.~\eqref{gsnrgV}. The resulting critical curves, shown in Fig. \ref{FmodinstEx2} are almost overlapping with the exact ones in Fig. \ref{FmodinstEx}.  
\section{Conclusions}
\label{conc}
In this paper we address the stability of stationary superfluid flow in the quantum phase model, adopting several mean-field schemes based on factorized trial states like in Eq. \eqref{prodTS}.
The Gaussian factors in Eq. \eqref{ansatzG} allow us to explicitly derive the equations of motion for the four macroscopic dynamical variables describing each lattice site, and to study analytically the stability of the attendant current-carrying stationary states. The critical lines for the dynamical instability of currents flowing along a coordinate direction are derived for arbitrary lattice dimensionality, Eq. \eqref{Gmodinst} and Fig. \ref{Fmodinst}. We also analytically prove that the current attains its maximum value along these curves, as argued and shown numerically for the (mean-field) Bose-Hubbard model \cite{Altman_PRL_95_020402,Polkovnikov_PRA_71_063613}.
Furthermore, we show that, energetic instability coincides with dynamical instability for the quantum phase model. As we discuss, this result is consistent with other models of interacting lattice bosons, such as the discrete nonlinear Schr\"odinger equations or the Bose-Hubbard model. 
In Sec. \ref{numres} we compare the results obtained from the  Gaussian factors in Eq. \eqref{ansatzG} with those obtained numerically in the standard mean-field description corresponding to the unconstrained factors in Eq. \eqref{TSfactorE}. Finally, in Sec. \ref{manres} we examine two further analytically tractable choices for the factors of Eq. \eqref{prodTS}, and show that the one-parameter choice in Eq. \eqref{Nansatz} provides analytical results that are in remarkably satisfactory agreement with those obtained numerically in the ``exact" mean-field scheme of Sec. \ref{numres}.
A study of the dynamics for such a trial state along the lines of Section \ref{anres} is currently underway \cite{Gutz_inprep}. This is expected to provide phase diagrams for the dynamical instability that are in  agreement with those obtained from the maximum current, shown in Fig.~\ref{FmodinstEx2}. Furthermore, since this description applies all the way to the (mean-field) critical point, it is expected to be useful in the study of the amplitude ``Higgs-like" excitations on two-dimensional lattices \cite{Huber_PRL_100_050404,Endres_Nature_487_454,Pollet_PRL_109_010401}. 

\acknowledgments
We thank E. Ercolessi for fruitful discussions.
This work has been supported by the EU-STREP Project QIBEC.
QSTAR is the MPQ, IIT, LENS, UniFi joint center for Quantum Science
and Technology in Arcetri.

\appendix
\section{Equations of motion}
\label{Atdvp}
Equations \eqref{eqmot} are obtained through the so-called time-dependent variational principle (TDVP), in which the system state is assumed to have the form $|\Psi\rangle= e^{i{\cal S}(t)} |\Xi \rangle$, where $ |\Xi \rangle$  is a trial state depending on a set of microscopic dynamical variables \cite{Amico_PRL_80_2189}. A careful choice $|\Xi\rangle$ and of the microscopic variables thereof, along with the requirement that the state of the system satisfies a weak form of the Schr\"odinger equation $\langle \Psi| i \partial_t - H |\Psi \rangle=0$ allows the identification of ${\cal S}$ as an effective action for the dynamical variables 
\begin{equation}
\dot {\cal S} = {\cal L}[\Xi]= \langle \Xi | i \partial_t |\Xi \rangle -  \langle \Xi | H |\Xi \rangle
\end{equation}
provided that $ {\cal L}[\Xi]$ and  
\begin{equation}
{\cal H}[\Xi]= \langle \Xi | H |\Xi \rangle
\end{equation}
are consistently recognized as the effective Lagrangian and Hamiltonian, respectively. The above notation emphasizes that these are functions of the set of microscopic variables  characterizing the state $|\Xi \rangle$, which we denote $\Xi=\{\xi_j\}$. The equations of motion for the latter therefore ensue from the stationarization of the action $\delta {\cal S}$, i.e. from the Euler-Lagrange equations
\begin{equation}
\left(\frac{d}{d t} \frac{\partial }{\partial \dot \xi_j}-\frac{\partial }{\partial  \xi_j}\right){\cal L}[\Xi]= 0
\end{equation}
 or, from the corresponding Hamilton equations. 
 
Time-dependent variational principles based on several different choices for the trial state $|\Xi \rangle$ have been e.g. applied to the Bose-Hubbard model. The resulting equations of motion are the well-known discrete-nonlinear Schr\"odinger equations if the trial state is either a product  Glauber coherent states, one for each of the $L$ lattice sites \cite{Amico_PRL_80_2189}, or a suitable SU($L$) coherent state \cite{Buonsante_PRA_72_043620}. If instead $|\Xi \rangle$ is assumed to have the form in Eq. \eqref{prodTS}, where each factor $|\psi_\mathbf{r}\rangle$ is a generic superposition of on-site Fock states -- as in Eq. \eqref{TSfactorE} -- the so-called time-dependent Gutzwiller equations are obtained \cite{Jaksch_PRL_89_040402}.

In the ansatz defined by Eqs. \eqref{prodTS} and \eqref{ansatzG} the trial state is a product of factors, each referring to a lattice site $\mathbf{r}$ and depending on four dynamical variables, $\varphi_\mathbf{r}$, $\sigma_\mathbf{r}$, $\varepsilon_\mathbf{r}$ and $\kappa_\mathbf{r}$.  
Straightforward if cumbersome calculations show that 
\begin{equation}
{\cal L} = \sum_\mathbf{r} \left(\kappa_\mathbf{r} {\dot \varphi}_\mathbf{r}-\sigma_\mathbf{r} {\dot \varepsilon}_\mathbf{r} \right)-{\cal H}
\end{equation}
where $\cal H$ has the form in Eq. \eqref{mfHG}, which confirms that $\varphi_\mathbf{r}$ and $\varepsilon_\mathbf{r}$ are conjugate to $\kappa_\mathbf{r}$ and $\sigma_\mathbf{r}$, respectively, and that their dynamics is dictated by Eq. \eqref{eqmot}.
\section{Bogoliubov frequencies}
\label{BogFreq}
The linear stability character of the (current-carrying) stationary states  is encoded in the spectrum of the matrix $\Omega$ governing the dynamics of the small deviations $\mathbf{y}_\mathbf{r}$ from the values in Eq. \eqref{ccss}
\begin{equation}
\label{deviat}
\varphi_\mathbf{r} = p\, x+ y_\mathbf{r}^{(1)},\; \kappa_\mathbf{r} = y_\mathbf{r}^{(2)},\;
\sigma_\mathbf{r} = \bar \sigma+ y_\mathbf{r}^{(3)},\;\varepsilon_\mathbf{r} = y_\mathbf{r}^{(4)}
\end{equation}
Plugging Eq. \eqref{deviat} into Eq. \eqref{eqmot} and retaining only the linear terms in $\mathbf{y}_\mathbf{r}$ one ends up with a linear equation $\dot{\mathbf{y}}_\mathbf{r} = i \Omega \mathbf{y}_\mathbf{r}$. The rank of $\Omega$ is $4 L$, where $L = \prod_{j=1}^d L_j$ is the number of sites in the lattice. Owing to translation invariance, switching to the reciprocal lattice decouples the problem into $L$ linear problems of rank $4$,    $\dot{\mathbf{y}}_\mathbf{q} = i \Omega_\mathbf{q} \mathbf{y}_\mathbf{q}$, with 
\begin{equation}
\label{FT}
\mathbf{y}_\mathbf{q} =(L)^{-\frac{1}{2}} \sum_\mathbf{r} e^{-i \mathbf{q}\cdot \mathbf{r}}\mathbf{y}_\mathbf{q}.
\end{equation}
 The characteristic polynomial of $\Omega_\mathbf{q}$ has the form in Eq. \eqref{char_pol}, where 
\begin{widetext}
\begin{align}
\label{cp_b}
b(p,\mathbf{q}) &=   2 e^{-2\bar \sigma} d\,\Gamma  \left[ (2 \bar \sigma -5) C_1+\left(2 \bar\sigma +1\right)   C_2\right]\\
\label{cp_c}
 c(p,\mathbf{q}) & =
32 e^{-4 \bar \sigma^2} d^2\, \Gamma^2\left\{  (C_1-C_2)\left[C_1-\bar \sigma\,C_2+(1-\bar \sigma)C_1\right]-\bar \sigma \sin^2 p \, \sin^2 q_1\right\}
\end{align}
\end{widetext}
and 
\begin{align}
\label{C1}
C_1 &= d-1+\cos p \\
\label{C2}
C_2 &= \cos p \cos q_1 + \sum_{j=2}^{d} \cos q_j
\end{align}
The conditions for linear stability, Eqs. \eqref{stability}, can be easily studied numerically, and shown to agree with the conclusions in Sec. \ref{anres}.
The analytical study of Eqs. \eqref{stability}  is sraightforward for one-dimensional lattices, but rather lengthy and tedious in the general case. Here we limit ourselves to sketching the study of the condition actually resulting in the critical line of Eq. \eqref{Gmodinst}, namely $c>0$.  
When $\bar \sigma>1$ there is a choice of $\mathbf{q}$ making $c$ negative irrespective of $p$.
Note indeed that, for $d>1$, $C_1\geq 0$ and $-C_1\leq C_2 \leq C_1$. In particular $C_2 \to  C_1$ when $\mathbf{q} \to \mathbf{0}$. Thus the sign of the product in the curly brackets of Eq. \eqref{cp_c} is determined by the factor in the square brackets, which is clearly negative for vanishing $\mathbf{q}$. Our claim is proven after noticing that  the remaining term in the curly brackets is negative irrespective of $\mathbf{q}$ and $p$.

A more detailed argument is needed to recognize the relevance of the threshold  in Eq. \eqref{Gmodinst} for   the  case $\bar \sigma <1$.
The coefficient $c$ in Eq. \eqref{cp_c} is an ``upward" paraboloid in the variables $\cos q_j$, and, in general, it is positive outside a $d-1$ ellipsoidal surface. Since $1 \leq \cos q_j \leq 1$, the stationary state becomes unstable as soon as this ellipsoidal surface intersects the hypecube of edge $2$ centered at the origin of axes (it can be checked that the center of the ellipsoid lies outside such hypercube). It turns out that the intersection occurs at the hypercube edge corresponding to $q_2 = q_3 = \cdots = q_d = 0$, where 
\begin{widetext}
\begin{equation}
c(p,\mathbf{q}) = \bar \sigma\, (\cos q_1-1)\left\{\cos q_1 +1 - \frac{2 \cos p}{\bar \sigma} \left(d-1+\cos p\right)-2 [1+(d-1) \cos p]\right\}
\end{equation}
\end{widetext}
Recalling that $-1\leq\cos q_1 <1$, the stability condition $c>0$ can be recast in terms of the nontrivial root of $c(p,\mathbf{q})$, namely the one in curly brackets, as 
\begin{equation}
2\frac{ (d-1+\cos p)\cos p -\sigma [1+(d-1) \cos p]}{\sigma}-1>1
\end{equation}
which is easily shown to be equivalent to the  condition in Eq. \eqref{Fd}.
\section{Energy instability}
\label{EnFreq}
The procedure for determining the energy stability character of  stationary states is similar to that illustrated in the previous section. Once again, a perturbation of the stationary state like in Eq. \eqref{deviat} is considered. Plugging it in Eq. \eqref{mfHG} and considering terms up to the second order in the perturbations,  one gets
\begin{equation}
\label{pertH}
{\cal H} = E_p + \sum_{\mathbf q} \mathbf{y}_{\mathbf q}^{\rm t} \cdot \Lambda_{\mathbf q}  \, \mathbf{y}_{\mathbf q}
\end{equation}
where once again we made use of Eq. \eqref{FT} and the problem decoupled owing to translation invariance. The  $4\times 4 $ matrix in Eq. \eqref{pertH} is related to $\Omega_\mathbf{q}$ as
\begin{equation}
\label{OmeLam}
\Omega_\mathbf{q} = -2
\left(\begin{array}{cc}
\sigma_y & 0\\
0 & \sigma_y
\end{array}\right) 
\Lambda_\mathbf{q}
\end{equation}
where $\sigma_y$ here denotes a Pauli matrix. Straightforward calculations show that the characteristic polynomial has the form in Eq. \eqref{char_pol_E}, with
\begin{equation}
\label{bcnrg}
b'(p,\mathbf{q}) = - \frac{e^{-2 \bar \sigma}}{\bar \sigma} \left(C_1 - \bar \sigma \,C_2 \right),\;\; 
c'(p,\mathbf{q}) = \frac{c(p,\mathbf{q})}{32d^2 \Gamma^2 \sigma }
\end{equation}
where the quantities appearing in Eqs. \eqref{bcnrg} have been defined in Eqs. \eqref{cp_c}-\eqref{C2}. Now, as we discuss in  Appendix \ref{BogFreq}, $|C_2|\leq C_1$, so that for  $\bar \sigma >1$
there is always some $\mathbf{q}$ such that $b'(p,\mathbf{q})>0$. This means that, irrespective of  
$c'(p,\mathbf{q})$, a stationary state with $\bar \sigma >1$ is always energetically unstable. This is expected since in   Appendix \ref{BogFreq} we showed that such stationary states were dynamically unstable. If, conversely, $\bar \sigma <1$, $b'(p,\mathbf{q})$ is always non negative, and the energetic stability of the stationary state is encoded in $c'(p,\mathbf{q})$. But, as we notice in Eq. \eqref{bcnrg}, $c'(p,\mathbf{q})$ has the same sign as $c(p,\mathbf{q})$, and hence the threshold for dynamic stability also marks the boundary between energetically stable and unstable states.
We once again remark that this is a specific feature of the quantum phase model, and does not apply e.g. for the discrete nonlinear Schr\"odinger equation \cite{Smerzi_PRL_89_170402,Menotti_NJP_5_112} or the Gutzwiller approach to the Bose-Hubbard model \cite{Saito_PRA_86_023623,Gutz_inprep}.
%


\end{document}